# Fuzzy Logic based Autonomous Parking Systems - Part III: A Fuzzy Decision Tree System

Yu Wang and Xiaoxi Zhu

*Abstract—* **This paper proposes a robust design of Hybrid Fuzzy Controller for speed and steering angle control in an Intelligent Autonomous Parking System (IAPS). The Hybrid Fuzzy Controller consists of a Base Fuzzy Controller (BFC) and a Supervisory Fuzzy Decision Tree Controller (SFDTC). The BFC evolves from previous work on fuzzy logic control for unmanned parking and it ensures that optimal parking trajectory is achieved with minimal computational cost. SFDTC further increases the system robustness when there is noise in the operating environment. The design of SFDTC combines Decision Tree theory and fuzzy inference mechanism. A data training process is also formulated to achieve better control performance. As a result, IAPS equipped with the new Hybrid Fuzzy Controller with Fuzzy Decision Tree (HFC-FDT) demonstrates optimal performance with faster convergence and minimal deviation from optimal parking trajectory. The detailed design of Supervisory Fuzzy Decision Tree Controller is presented in this paper with a MATLAB simulated experiment which concludes the superior performance of the new design.**

## I. INTRODUCTION

Intelligent Transportation System (ITS) has aroused great research interest in the past few decades. Multiple research tracks have been identified in this area, among which autonomous parking has attracted most attention. Researchers in the area of control theory and Artificial Intelligence (AI) have been working on designing robust controllers for efficient unmanned parking. However, traditional and even advanced control algorithms with sophisticated mathematical models do not perform well in achieving optimal parking trajectory. One inherent obstacle is that parameters change fast during the parking process and unknown disturbances may arise from time to time. It is challenging to derive an accurate modelling and an optimal controller. Another drawback is the huge amount of on-line data processing required to implement such controllers. On a separate track, fuzzy logic control which uses symbolic modelling as opposed to mathematical modelling in conventional control theory has been widely explored.

Fuzzy logic control has been applied in various areas since it was first proposed by Zadeh in 1965 ([1],[2]). This technique has also been widely used to conquer challenges in unmanned transportation system such as speed control and adaptive cruise control ([3-5]). Recently an Intelligent Autonomous Parking System with a Fuzzy-Based Onboard System (FBOS) was proposed by Wang and Zhu in [6]. The proposed FBOS enables the system to achieve multiple functions - posture stabilization, steering angle control and path planning at minimal cost. More recent works also designed different algorithms based on Fuzzy logic control [7, 14-16]

In the development of Intelligent Autonomous Parking System, great research efforts have been put into the design of a best-performing controller for turning, which is the most critical step where even human beings find challenging as well. Two fuzzy logic controllers are designed in [6] to manipulate the unmanned vehicle in the process of turning - one to control the speed and the other to control the steering angle. Control decisions are made based on fuzzy inference that resembles human reasoning, which demonstrates satisfactory results. The intrinsic robustness of fuzzy logic controller ensures that the vehicle reaches the pre-set final state regardless of any noise. Another major advantage is that computation cost is reduced significantly compared to conventional controllers. However, the simple fuzzy logic controller does have disadvantages in terms of convergence and overshoot. Performance in these two areas deteriorates seriously when the noise level is high. In order to improve on this, a Hybrid Fuzzy Controller (HFC) is proposed in [7]. The HFC introduces a supervisory controller which fine tunes the control signal from Base Fuzzy Controller in situations where significantly large noise is present. Experiments and simulation results have proven the improvement in control performance as presented in [7].

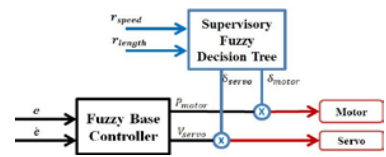

Fig. 1. Control Diagram of HFC-FDT

Following the discussion in [6] and [7], two important parameters that need to be monitored and controlled in the turning process are vehicle speed and steering angle, which are determined by the motor power of driving wheel and the servo voltage of driven wheel. It is also observed that two major types of noise that affect the whole parking process are ground condition and vehicle length. In previous work, it is assumed that the two factors can be decoupled, and adjustment is made by separate controllers. However, the

The corresponding author Yu Wang is with the Department of Electrical Engineering, Yale University, New Haven, CT, 06510, yu.wang@aya.yale.edu, Xiaoxi Zhu is with Google.

actual turning radius is determined by both vehicle speed and steering angle simultaneously. As a result, any deviation from the optimal trajectory is a joint effect of ground condition and vehicle length. As such, a revamped Hybrid Fuzzy Controller with Fuzzy Decision Tree (HFC-FDT) is proposed. The high-level control diagram is shown in Fig.1.

The underlying principle for Base Fuzzy Controller remain unchanged in the revised controller design. The key difference is that a single controller is used to control both actuators - motor (controlling vehicle speed) and servo (controlling steering angle) with two separate control signals. Inputs to the BFC are $e$ - difference between set reference and current position and $\dot{e}$ - change rate of this difference. Output control signal is derived using fuzzy inference based on Table I - *IF-THEN Rule for Base Fuzzy Controller*. Outputs of BFC are two control signals indicating the motor power rating $P_{motor}$ and servo voltage $V_{servo}$. In order to fine tune the output control signal from BFC to achieve better performance, a Supervisory Fuzzy Decision Tree Controller (SFDTC) is imposed to modify the BFC outputs. Inputs to SFDTC are measurements that represent the noise level in ground condition and vehicle length. Outputs are two scalars $\delta_{motor}$, $\delta_{servo}$ used to adjust BFC outputs. BFC outputs are multiplied by the corresponding scalar before sent to the actuator. With this approach, the impact of noise, i.e. variations in ground condition and vehicle length can be evaluated and corrected simultaneously.

| $\dot{e}$ \ $e$ | NL | NM | NS | ZO | PS | PM | PL |
|---|---|---|---|---|---|---|---|
| PL | NM | NS | NS | PS | PM | PL | PL |
| PS | NL | NM | NS | ZO | PM | PL | PL |
| ZO | NL | NM | NS | ZO | PS | PM | PL |
| NS | NL | NL | NM | ZO | PS | PM | PL |
| NL | NL | NL | NM | NS | PS | PS | PM |

TABLE I
IF-THEN RULE OF BASE FUZZY CONTROLLER

The theoretical foundation for the newly proposed SFDTC is fuzzy inference based Decision Tree. Decision Tree (DT) is a well-known algorithm widely used for fast classification and decision making. It has also been applied to ITS, in which case a satisfactory performance is achieved ([8]). Despite its advantage in symbolic domains, one of the shortcomings of DT lies with its weakness in handling uncertainties of numerical values ([9]). To overcome this issue, some researchers have proposed a new technique that combines fuzzy logic and decision tree, termed as Fuzzy Decision Trees (FDT). Theoretical analysis [10-13] has concluded that FDT should have better performance as it combines the advantages of fast classification of a Decision Tree and handling uncertainties of fuzzy theory. However, limited implementations have been carried out to demonstrate the validity, robustness and the efficiency of algorithm. Simulation results in Section III serve as a proof of the significant improvement in control performance with this advanced approach.

Another key advantage introduced by Supervisory Fuzzy Decision Tree Controller is the improvement of accuracy. In previous Supervisory Fuzzy Controller (SFC) design, the evaluation of noise level and selection of proper control signal is based on prior experience and experiment. With the proposed SFDTC, a formal training process of Decision Tree is deployed to derive the membership functions of input and output fuzzy variables. Detailed design and implementation of SFDTC is discussed in the next section.

## II. SUPERVISORY FUZZY DECISION TREE CONTROLLER

Following the discussion in the previous section, this section details the methodology used to construct the Supervisory Fuzzy Decision Tree Controller. The control goal is to fine tune the performance of FBOS while vehicle is turning, achieved by further adjustment of the two key control signals - motor power of the driving wheel and servo voltage of the driven wheel.

The SFDTC, as the name suggests, is constructed as a Decision Tree where the control decision is derived using fuzzy inference mechanism. Compared with the simple Supervisory Fuzzy Controller presented in [7], accuracy can be significantly improved in two ways:

1) control signal adjustments (for both driving wheel and driven wheels) are generated by calculating the negative impact of two types of noise simultaneously

2) numeric values of sample classification criteria and control signal are determined via rigorous training with large number of sample data.

More details of the proposed controller is discussed in the following three subsections:

1) Constructing Decision Tree - defining various components (root, node, leaf and path) to build the tree

2) Fuzzy Based Decision Making - applying fuzzy inference mechanism to decision making process

3) FDT training - training the decision tree with sample data to quantify the various components

### A. Constructing the Decision Tree

Decision Tree (DT) is one of the most important algorithms in discriminative learning, based on recursive partitioning. The objective is to classify data into different subsets by partitioning the sample set against evaluation criteria structured in different levels. Each partition creates one additional level in the tree structure. New data starts off from the root, evaluated against the partitioning criteria at each node and traverses down the path which represents the fulfilled condition until it reaches the leaf. Further actions can be applied based on the classification results.

The Decision Tree theorem is adopted in the controller design for Intelligent Autonomous Parking System. In conventional control theorem, controller is modelled as a mathematical function which processes the inputs and sends the calculated output(s) to the actuators. In contrast, a DT based controller assigns an individual control rule to each leaf node. New sample will be classified as belonging to a certain leaf node before the control rule associated with the leaf is executed.

As discussed in [6], two major factors that may deteriorate the control performance during turning are ground condition and vehicle length, a large variation in which will deviate vehicle speed and turning radius from optimal level. In this context, ground condition and vehicle length as inputs to SFDTC are not directly measured but defined as general parameters derived from other variables. These two parameters are also used as partition criteria in the Decision Tree.

The root node (level 0) of the Decision Tree $R_0$ contains a collection of elements, each represents a scenario where a fix-length vehicle is moving on a surface with certain ground conditions (either smooth or coarse). Each element at the root node is evaluated against the criteria "ground condition" and partitioned into three subsets "Coarse" ($N_1$), "Fair"($N_2$) and "Smooth" ($N_3$) corresponding to the three child nodes at next level. The path connecting $R_0$ and each child node is associated with a logic statement "Ground is coarse/fair/smooth". Each element traverses down the path where the logical statement is "True" based on the evaluation. The process is repeated for each level-1 node with the partition criteria "Vehicle length". The second partition generates in two child nodes for each level-1 node. As a result, there are six nodes at level-2, which are the leaf nodes - $L_{1,a}$, $L_{1,b}$, $L_{2,a}$, $L_{2,b}$, $L_{3,a}$, $L_{3,b}$.

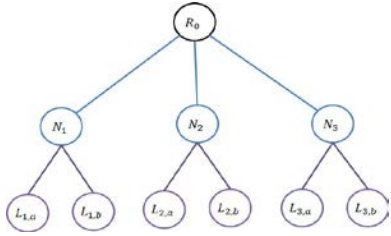

Fig. 2. Structure of Fuzzy Decision Tree

The partitioning ends at the leaf nodes and a particular control rules should be applied to elements inside each node. Each control rule contains two elements - motor power scalar which adjusts the speed of driving wheel and servo voltage scalar which adjusts the steering of driven wheel. Mathematically, the control rule can be represented by a vector with two elements [$\delta_{motor}$, $\delta_{servo}$].

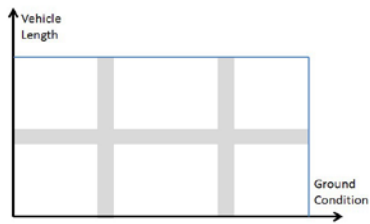

Fig. 3. Sample Space Partitioning by Fuzzy Decision Tree

The final Decision Tree is a three-level balanced tree presented in Fig.2. In accordance with the tree structure, the elements in the root node can be represented as a two-dimensional space partitioned into six subsets as in Fig.3. Each subset is equivalent to a leaf node. The shaded area between adjacent subsets indicates that the partitioning is based on linguistic variables instead of crisp values. The partitioning and decision making process is discussed in the next subsection.

### B. Fuzzy-based Decision Making

Conventional decision tree and partitioning is developed based on classical set theory, where a given element can either belongs to a set where $U(x) = 1$ or not where $U(x) = 0$. Given a node $N_l$ at level $l$, which has m child nodes at level $l+1$, thus $N_{l+1,1}$, $N_{l+1,2}$ $N_{l+1,1}$. If an element x belongs to the node $N_{l+1,k}$, it is defined such that $U(N_{l+1,k},x) = 1$. Otherwise $U(N_{l+1,k},x)=0$. Alternatively, since the path connecting $N_l$ and $N_{l+1,k}$ stands for the logic statement such that element x is $C_{l+1,k}$ where $C_{l+1,k}$ is the condition used to describe elements in node $N_{l+1,k}$, the logic statement is true if x belongs to the node. Each leaf node in a decision tree with $L+1$ levels can be defined by a compound logical statement: $P_1$ and $P_2$ and ...... $P_L$ where $P_k$ is the logical statement associated with each path that the element passes through before arriving at the leaf node. Classification of new element can also be viewed as mapping the data against compound logical statement at each leaf node until a match is found.

Decision tree with classical set theory has two inevitable disadvantages. First of all, the partition criteria shall be designed carefully to ensure comprehensibility and yet avoid overlapping. Besides, a well-defined threshold between two subsets is difficult to drawn if sample data used for training are clustered. To overcome the above mentioned constraints, a decision tree with fuzzy set theory is introduced.

*1) Node Partitioning with Fuzzy set theory:* In fuzzy set theory, the universe of discourse is divided into multiple subsets, where each subset can be identified by a linguistic variable with a membership function. In contrast to classical set theory where the membership function only has two discrete values {0,1}, the membership function of a linguistic variable is a continuous set on [0,1]. Taking a different angle on this, the logic statement for each path can be either truth or false in decision tree built using classical set theory. In the proposed fuzzy-based decision tree, however, the logic statement is assigned a level of truth which is a value between zero and one. Another major difference compared to classical set theory is that the membership functions of different fuzzy subsets can overlap with each other, which ensures complete coverage of the sample space. Such overlapping also indicates that a given element can belong to two subsets. Therefore, two logic statements defined on different paths can be true at the same time, but with a different degree of truthfulness. In this way, a transition region is generated and it resolves the issue where a crisp border line is difficult to define.

Take the partitioning at root node as an example to illustrate the fuzzy based decision making process. Each sample data shall be classified based on the partition criteria "ground condition" and ends up in any of child nodes at next level "Coarse", "Fair" and "Smooth". As the variable "ground condition" is continuous, the membership function based on classical set theory is shown in Fig.4, where the value can be either 1 or 0. With fuzzy set theory, the membership function of each linguistic variable can be modified as in Fig.5 with its value continuous on the interval [0,1]. The most commonly used membership function in fuzzy theory – "trapezoidal" - is proposed here. The element whose ground condition falls into the region between $\alpha_1$ and $\alpha_2$ can traverse down either of the two paths with different probability. The selection of $\alpha_1$ and $\alpha_2$ shall be further discussed in next subsection.

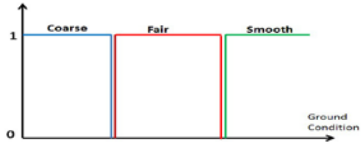

Fig. 4. Ground Conditions with Classical Set Theory

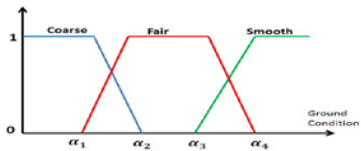

Fig. 5. Ground Conditions with Fuzzy Set Theory

The second partition will be based on vehicle length and two linguistic variables are defined - "long" and "short". As the partition is dependent on the previous partition, each node in level 1 generates a set of child nodes in the next level. Therefore, the membership functions defined on the child node can be different if they belong to different parent node even though the linguistic variable remains the same. The membership function for all six nodes in level-2 are illustrated in Fig.6.

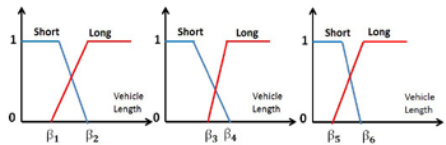

Fig. 6. vehicle Length with Fuzzy Set Theory

Previous discussion has concluded that a control decision - a vector consisting of two control signals - should be assigned to each leaf node . The control signals are also fuzzified with the membership functions shown in Fig.7. As a result of fuzzy based partitioning, each element may end up in multiple leaf nodes with different probabilities, which add up to unity. As a control decision is defined on each leaf node, an appropriate algorithm should be defined to obtain a single output control signal. The details of decision making process will be discussed in the next subsection.

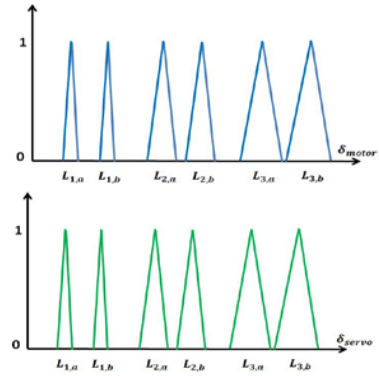

Fig. 7. Fuzzy Control Signals

2) *Decision making with fuzzy inference:* With the introduction of modified partitioning process based on fuzzy set theory, the decision making process shall be revised with fuzzy inference mechanism as well. The approach is explained with an example.

Define the sample data as a two-dimensional vector with elements' ground condition and vehicle length [grd, len]. Both elements of the input vector have a crisp value. The first step is fuzzification - to map the crisp value onto a respective fuzzy membership function. The mapping process is illustrated in Fig.8. As the compound logic statement of a particular leaf node is an intersection of two conditions, "min" operator is used in the fuzzy inference. The output linguistic variable is clipped at the minimum level of all fuzzy inputs.

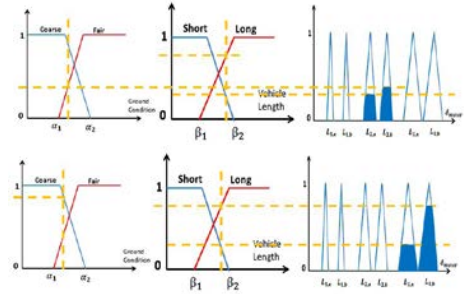

Fig. 8. Decision Making with Fuzzy Inference

Since each sample data may end up in multiple leaf nodes, the same input vector [grd,len] can lead to another output fuzzy set using the same fuzzy inference mechanism as shown in Fig.8. In reality, actuators cannot interpret fuzzy sets, thus only one single crisp control signal is allowed. Therefore, multiple fuzzy outputs are aggregated and de-fuzzified to derive the final control output. Defuzzification is achieved by deriving the centre of gravity (as illustrated in Fig.9).

*C. DFT trainging*

The previous two sections detail the supervisory controller design based on a combination of decision tree theory and fuzzy logic inference. The general structure of decision tree and decision making process is defined with prior knowledge. However, to achieve an optimal control performance, membership function of each linguistic variable, and numeric values of control signals, must be defined with a scientific approach. One big improvement compared to previous work is the formal data training process proposed here to derive the optimal partition criteria and control decisions.

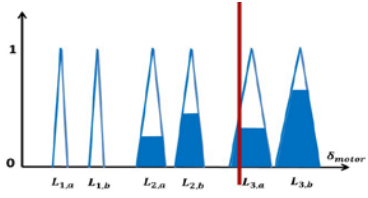

Fig. 9. De-fuzzification of Output Control Signal

In order to classify the ground condition as "Coarse", "Fair" and "Smooth", and the vehicle length as "Long" and "Short", a numeric value is needed for each variable.

As ground condition cannot be directly measured, a new indicator is defined to represent the ground condition. Set the motor power of driving wheel to a fixed value $P_{testing}$ and measure the speed under this driving power. Take the ratio $r_{speed}$ of the current speed against the optimal speed $v_{optimal}$ as the indicator of ground condition. The larger $r_{speed}$ is, the smoother the ground condition. $P_{testing}$ is selected such that under normal conditions, the vehicle should proceed with the $v_{optimal}$ which is the optimal speed for the parking process.

Although vehicle length can be obtained directly, there are other factors that may affect the turning radius as well such as position of the wheels. Therefore, the ratio $r_{length}$ of actual turning radius versus the optimal turning radius is taken as the indicator of vehicle length. Larger $r_{length}$ indicates a longer vehicle. The actual turning ratio is calculated by dividing vehicle speed by angular velocity measured from Inertial Measurement Unit (IMU). The angular velocity shall be measured when the vehicle is turning at a constant speed $v_{optimal}$ and the servo voltage of driven wheel is set at a fixed value $V_{testing}$.

The following steps are taken to collect data used for Decision Tree training.

1) Set the power rating of motor to $P_{testing}$ and measure the vehicle speed. Take down the the ratio $r_{speed}$ of the measured speed against the optimal speed as an indicator of ground condition.

2) Slowly tune the motor power (increase if $r_{speed}$ is greater than one and vice versa) until the vehicle speed reaches the optimal level $v_{optimal}$. Once the speed is steady, steer the driven wheel by applying the testing servo voltage $V_{testing}$ and the vehicle will start turning. The angular velocity can be measured by IMU. Obtain the current turning radius by dividing speed by angular velocity. Take the ratio $r_{length}$ of current turning radius and the optimal value as an indicator of vehicle length.

3) Slowly tune the servo voltage and motor power in opposite direction till the turning radius reaches the desired level. Record down another two ratios: the final values of motor power rating against $P_{testing}$ as $\delta_{power}$ and servo voltage against the testing value $V_{testing}$ as $\delta_{servo}$. Repeat the experiment for different types of vehicles under various ground conditions. Plot the two indicators ($r_{speed}$ for ground condition and $r_{length}$ for vehicle length) in a two dimensional space as shown in Fig.10.

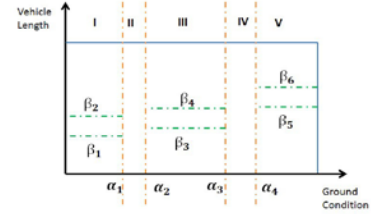

Fig. 10. Space Partitioning with FDT Training

The horizontal axis can be divided into five regions with four vertical lines, which generates the membership functions of each linguistic variable defined for ground condition. The membership function of respective linguistic variable "coarse", "fair" and "smooth" corresponding to each region (I, III, V) is defined as one while the rest as zero. Region II and IV are viewed as transition regions where two linguistic variables have non-zero membership function at the same time. The membership functions is shown in Fig.5.

Each of the regions I, III, V can be further divided into three segments by two horizontal lines, illustrating partition of "long" and "short" vehicles. The top or bottom segment in each region represents "long" or "short" vehicles. Membership function of the linguistic variable corresponding to individual segment is one. The middle segment is the transition segment where both linguistic variables have non-zero membership function values and add up to one. It shall be noted that the segment partitioning can be different for each region (I, III and V). The membership functions for linguistic variable "vehicle length" can be generated accordingly (shown in Fig.6)

As a result, six major categories are identified based on the ground condition and vehicle length measurements. Each of the categories corresponds to a leaf node in the decision tree. A control rule shall be defined for each of the category or leaf node. The measurements $\delta_{motor}$ and obtained in data training can be used as reference to design the control rules. Plot the value of $\delta_{motor}$ as z-axis for each sample data that falls into the six major categories. The median value for each category shall be used as the control signal (as shown in Fig.11). Apply the same method to obtain the value of servo voltage scalar $\delta_{servo}$.

III. EXPERIMENT RESULTS

The proposed Hybrid Fuzzy Controller with Fuzzy Decision Tree (HFC-FDT) is simulated in MATLAB to validate the performance improvement compared with Fuzzy-Based Onboard System (FBOS) and Hybrid Fuzzy Controller (HFC) in previous work. Parallel experiments are simulated for autonomous parking process with same disturbances - large friction ground condition and long vehicle, controlled by different types of controllers FBOS, HFC and HFC-FDT.

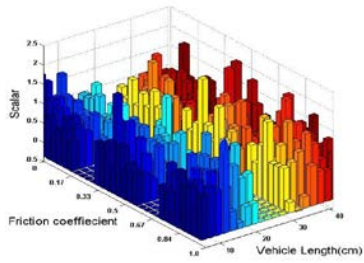

Fig. 11. Determine Control Signal value with FDT Training

All three controllers need to be configured beforehand. HFC-FDT is configured following the data training process elaborated in Section II.C. The parameters in FBOS and HFC are also configured accordingly based on trial experiments.

By recording and tracking the path of the entire parallel parking process, a straightforward comparison of the performance of different controllers can be obtained. The interval time t between each two recordings is a constant, which is 0.1 second. By fixing the time interval across different controllers, efficiency of the entire parking process can be easily compared. The more scattered the tracking figures, the faster of the process. Fig.12 shows a comparison among three parallel parking process by implementing FBOS, HFC and HFC-FDT (from left to right). It is observed that both cars controlled by HFC and HFC-FDT can be parked properly with a satisfactory parking trajectory. The car with basic FBOS, however, ends up colliding into the neighbouring slot due to a significant path distortion . This is mainly because that large friction and vehicle size impede the steering of driven wheels and the rotation of driving wheels. Hence the parking trajectory is distorted badly. A further comparison of the trajectories of vehicles controlled by HFC and HFC-FDT shows a difference in efficiency. Plot of vehicle controlled by HFC-FDT shows more scattered dots and smoother turning trajectory. On the other hand, vehicle controlled by HFC has a longer transient period and denser plot. The difference proves that the vehicle controlled by HFC-FDT is operating with higher efficiency. In conclusion, the improved HFC-FDT gives superior control performance in terms of accuracy and efficiency.

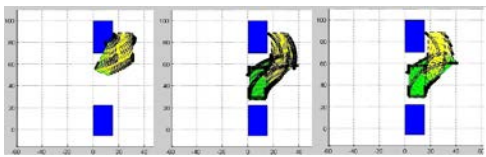

Fig. 12. Performance Comparison of Different Controllers

## IV. CONCLUSIONS AND FUTURE WORK

In this paper, a Hybrid Fuzzy Controller with Fuzzy Decision Tree is proposed. The Supervisory Fuzzy Decision Tree Controller significantly improves the control performance compared to the original FBOS and enhanced HFC in previous work ([6],[7]). In SFDTC, a three-level decision tree is constructed to classify all sample data into six major categories based on two types of noise in the parking process - ground condition and vehicle length. Each category corresponds to a leaf node in the decision tree and an appropriate control rule is applied to the leaf node. The decision tree is trained with large amount of sample data to generate the partition criteria and control rule. The decision making process with trained decision tree is based on fuzzy inference which further improves the robustness of the controller. However, one drawback in this approach is increased efforts required in decision tree training.

Decision tree is one of the most fundamental data classification techniques. One interesting area for future study is to compare the performance of controllers built with other classification methodologies, such as neural network and cooperative game theory. However, one trade-off associated with more advanced theory is the increase in computation power required. The drawback must be carefully weighed against the improvement in performance.